\documentclass[10pt]{article}

\usepackage[letterpaper,top=0.70in,bottom=0.74in,left=0.78in,right=0.78in]{geometry}
\usepackage[T1]{fontenc}
\usepackage{amsmath,amssymb,mathtools,amsthm}
\usepackage{newtxtext,newtxmath}
\usepackage{microtype}
\usepackage{booktabs,tabularx,array,multirow}
\usepackage{enumitem}
\usepackage[dvipsnames]{xcolor}
\usepackage{graphicx}
\usepackage{tikz}
\usetikzlibrary{arrows.meta,positioning,fit,calc,shapes.geometric,backgrounds}
\usepackage[font=small,labelfont=bf]{caption}
\usepackage[numbers,sort&compress]{natbib}
\usepackage{fancyhdr}
\usepackage{titlesec}
\usepackage{balance}
\usepackage{url}
\usepackage[colorlinks=true,allcolors=NavyBlue]{hyperref}
\usepackage[capitalise,noabbrev,nameinlink]{cleveref}

\setlist{nosep,leftmargin=1.4em}
\titleformat{\section}{\large\bfseries}{\thesection}{0.55em}{}
\titleformat{\subsection}{\normalsize\bfseries}{\thesubsection}{0.5em}{}
\titleformat{\paragraph}[runin]{\bfseries}{\theparagraph}{0.45em}{}[.]
\titlespacing*{\section}{0pt}{1.0em}{.40em}
\titlespacing*{\subsection}{0pt}{.76em}{.27em}

\newtheorem{proposition}{Proposition}
\newtheorem{definition}{Definition}
\newcommand{\Org}{\mathcal{O}}
\newcommand{\Agents}{\mathcal{A}}
\newcommand{\Events}{\mathcal{E}}
\newcommand{\Runtime}{\mathcal{R}}
\newcommand{\Artifacts}{\mathcal{F}}
\newcommand{\Knowledge}{\mathcal{K}}
\newcommand{\Permission}{\mathsf{P}}
\newcommand{\Privilege}{\mathsf{Q}}
\newcolumntype{Y}{>{\raggedright\arraybackslash}X}

\title{\vspace{-1.2em}\textbf{Fluid Structure, Rigid Record:}\\
\large A Layered Organizational Design Framework for Agent-Native Organizations}
\author{Lucian Zhu\\\small Independent Researcher\\\small\texttt{Business@thisislucian.com}}
\date{August 2026 -- Independent Draft v2}

\begin{document}
\maketitle
\vspace{-1.2em}

\begin{abstract}
An organization is not a set of model instances with corporate titles; it is a goal-directed system that turns specialization into coordinated output. Most large-language-model multi-agent systems still operationalize organization as a conversational topology, a role prompt, or a fixed workflow. This paper develops an organizationally structured design framework in which the stable and changing parts are separated. The persistent layer contains a four-store record system and a pool of resident specialization templates. A coordination layer uses \emph{Permission}---the boundary of information, tools, and channels an agent can access---and \emph{Privilege}---the organizational state changes it may authorize---to compile a task-specific world. A runtime layer combines an external Workflow Protocol, an isolated runtime store, and a dynamically assembled Task Group. A human-interaction layer exposes the system through a Control Plane and a non-decision-making Translation Agent. Three orthogonal Role Groups then separate Operation, Review, and Supervision: operators execute inside narrow leases; reviewers hold high but on-demand change authority; supervisors receive broad observational access but little modifying authority. The result is fluid at the surface and rigid underneath: events and tasks change team composition, topology, views, tools, and workflow, while records, write rules, and separation of powers persist. The architecture has been implemented as a prototype and exercised in small-sample tests; large-scale empirical validation is not yet complete, so no broad performance claim is made. The contribution is a coherent, falsifiable framework for designing, governing, recovering, and evaluating agent organizations.
\end{abstract}


\section{Introduction}

Organizations exist because differentiated workers can achieve a shared objective more efficiently together than separately. The operative words are \emph{objective}, \emph{differentiation}, and \emph{coordination}. A collection of agents is therefore not an organization merely because its prompts say ``CEO,'' ``analyst,'' or ``reviewer.'' It becomes one only when it can select capabilities for a goal, manage dependencies, control information and action boundaries, transfer evidence, absorb failure, and preserve an accountable history of work.

Current LLM multi-agent systems provide important pieces of this problem. CAMEL and AutoGen make conversational roles programmable \citep{li2023camel,wu2023autogen}; MetaGPT couples specialist roles to a software workflow \citep{hong2023metagpt}; MultiAgentBench studies coordination and competition \citep{zhu2025multiagentbench}; and OneManCompany (OMC) introduces portable talents, dynamic recruitment, and typed organizational interfaces \citep{yu2026skills}. These advances should be credited, but they do not make every organizational concern first-class. In particular, durable work can remain entangled with sessions, authority can remain implicit in tool availability, review can be a prompt rather than a constrained power, and recovery can mean asking a contaminated context to try again.

The gap is consequential. MAST locates common failures in specification and system design, inter-agent alignment, and verification or termination across 1,642 execution traces \citep{cemri2025mast}. Controlled scaling work further shows that additional agents are not monotonically beneficial: centralized coordination can help genuinely parallel work, whereas matched-budget multi-agent configurations can sharply degrade sequential reasoning and amplify unchecked errors \citep{kim2025scaling}. Tool security adds another warning: when untrusted text and powerful actions share a channel, prompt-level guidance is not an adequate authorization boundary \citep{debenedetti2024agentdojo,greshake2023injection}.

This paper takes an organization-design rather than a model-centric view. Its core proposition is:

\begin{quote}
\centering\itshape
Change Permissions and Privileges to drive work, not prompts. Compile the world an agent may inhabit, then let it act freely inside that world.
\end{quote}

\textbf{Permission} determines what world an agent can observe and touch: documents, tools, channels, and writable regions. \textbf{Privilege} determines what changes it may make to the organization: approve, promote, spawn, halt, grant, publish, or alter policy. Prompts can propose; only externally enforced P\&P can authorize. These two variables connect two persistent modules---the record system and the Specialization/Role-Group Agent Pool---to a temporary execution surface assembled by the Workflow Protocol. This composition, rather than any single component, is the paper's central comparative advantage.

The resulting architecture is \emph{fluid where work varies and rigid where accountability must survive}. There is no digital employee permanently nailed to a position. An event or task causes the system to select resident specializations, instantiate temporary workers, choose a task topology, cache an authorized working set, and issue expiring P\&P. When work ends, the team dissolves and the grants disappear. What persists is the specialization template, the separation of Role Groups, the record, and the rules by which organizational state may change.

The paper makes four contributions:
\begin{enumerate}
  \item a functional translation method that imports mechanisms from organization and management research by the problem they solve, not by their human surface form;
  \item a four-layer reference architecture joining a rigid record system, a specialization pool, P\&P coordination, dynamic Task Groups, and a human Control Plane;
  \item a formal separation of three Role Groups whose Permission and Privilege profiles are deliberately asymmetric; and
  \item an integrated failure-protection and measurement loop with scoped restart, replay, escalation, circuit breaking, and independent shutdown.
\end{enumerate}

This is the first paper in a planned sequence. It concentrates on the framework and its internal logic. A companion paper will report controlled, larger-sample evaluation; subsequent work will examine scale and organizational self-evolution. The present prototype and small-sample exercises establish implementability, not general performance.

\section{An Organization-Design Lens}

\subsection{Translate functions, not corporate theatre}

Classical organization research is relevant because human and artificial collectives share some functional constraints: differentiated capability, interdependence, limited attention, uncertainty, information aggregation, and error control. They do not share every human constraint. Employment, status, morale, career, tacit responsibility, and political legitimacy do not automatically transfer to model instances. The appropriate method is therefore selective translation.

\begin{definition}[Functional translation criterion]
A mechanism from human organization design is transferred only after identifying the failure-generating constraint it addresses. If the constraint exists for agent systems, the mechanism is reimplemented in machine-enforceable form. If it does not, the surface form is deleted. If the evidence is incomplete, the mechanism remains a testable design hypothesis rather than an invariant.
\end{definition}

This criterion turns organization studies into design material rather than metaphor. \Cref{tab:theory} shows the main translations used here. The distinctive research niche is not that management language is applied to MAS; recent work already argues for machine-compatible organizational principles \citep{xian2025reliable}. It is that differentiation, integration, authority, records, and recovery are composed into one operational framework.

\begin{table}[t]
\centering
\caption{Organization-theory mechanisms translated into machine-enforceable design.}
\label{tab:theory}
\footnotesize
\begin{tabularx}{\textwidth}{@{}p{3.0cm}YY@{}}
\toprule
Source mechanism & Functional problem & Translation in this framework \\
\midrule
Bounded rationality and attention \citep{march1958organizations} & no actor can process the whole organization at once & Permission compiles a narrow task view; context is a controlled resource \\
Differentiation and integration \citep{lawrence1967organization} & specialization creates value but also coordination demand & a persistent specialization pool is integrated per task by P\&P and workflow \\
Information-processing design \citep{galbraith1974organization} & uncertainty raises information-processing requirements & views, evidence, review depth, and aggregation points expand with task uncertainty \\
Interdependence types \citep{thompson1967organizations} & pooled, sequential, and reciprocal work require different coordination & topology follows the dependency graph rather than a fixed org chart \\
Mechanistic/organic contingency \citep{burns1961innovation} & no single structure fits stable and changing conditions & persistent rules remain rigid while runtime structure is fluid \\
Enabling formalization \citep{adler1996bureaucracy} & rules can support repair and visibility rather than merely coerce & typed cards, state, lineage, and error paths help agents act and recover inside boundaries \\
Coordination theory \citep{malone1994coordination} & coordination is management of dependencies among activities & the Workflow Protocol manages dependencies; messages do not carry canonical state \\
High-reliability defense \citep{reason1997managing} & independent safeguards must prevent aligned failure paths & prevention, detection, containment, recovery, and learning are separated \\
\bottomrule
\end{tabularx}
\end{table}

This lens also rejects a false choice between hierarchy and flatness. Status hierarchy is unnecessary; information aggregation is not. A sequential task may require only one chain. A parallel task may need an aggregation tree because a synthesizer has finite context. The topology is justified by the information problem, not inherited from human rank.

\subsection{Four design laws}

The framework is constrained by four laws.

\paragraph{L1: Dormancy is the default.}
An agent without an active task has zero standing organizational P\&P beyond a private workspace and harmless baseline functions. A task activates a scoped lease; completion, timeout, budget break, or HALT revokes it. The normal state is closed, so failure tends toward safety rather than residual autonomy.

\paragraph{L2: Specialization justifies residency.}
A resident agent specification exists only because a durable capability bundle is repeatedly useful: domain context, tools, skills, evaluation priors, or a stable working style. Residency is never justified by a corporate title. Commodity work is temporary.

\paragraph{L3: An organizational identity is specialization times Role Group.}
A complete specification is
\begin{equation}
  \operatorname{AgentSpec}=\operatorname{Specialization}\times\operatorname{RoleGroup},
  \label{eq:agentspec}
\end{equation}
where specialization describes what the agent is equipped to do and Role Group defines the shape of its P\&P. Experience may improve the specialization, but it does not create rank-based authority.

\paragraph{L4: Instantiation and topology follow the task.}
A resident specialist is best understood as a reusable team lead or template. It may instantiate temporary, possibly heterogeneous workers that inherit attenuated P\&P and cannot expand their own scope. The dependency shape then determines whether these workers form one chain, independent branches, an aggregation tree, or a hybrid. Fan-out is explicit rather than allowed to grow accidentally.

Three corollaries recur throughout the paper: \emph{write work as records and pass pointers, not sessions}; \emph{no agent controls every power}; and \emph{manage the world rather than narrate behavior inside a noisy world}. These are not stylistic preferences. They specify where enforceable coordination lives.
\section{The Four-Layer Architecture}

\subsection{Integrated view}

At time $t$, define the organization as
\begin{equation}
\Org_t=\left\langle \mathcal{L}_{\mathrm{per}},\mathcal{L}_{\mathrm{coord}},
\mathcal{L}_{\mathrm{run}},\mathcal{L}_{\mathrm{human}},\Pi\right\rangle,
\label{eq:organization}
\end{equation}
where $\Pi$ is policy enforced outside model inference. The four layers are:
\begin{align}
\mathcal{L}_{\mathrm{per}} &= \langle \Agents^{\mathrm{res}},\Events,\Artifacts,\Knowledge\rangle,\\
\mathcal{L}_{\mathrm{coord}} &= \langle \Permission_t,\Privilege_t,\Gamma_t\rangle,\\
\mathcal{L}_{\mathrm{run}} &= \langle \Runtime_t,W,G_t\rangle,\\
\mathcal{L}_{\mathrm{human}} &= \langle H,C,T\rangle.
\end{align}
Here $\Agents^{\mathrm{res}}$ is the resident specialization pool; $\Events$, $\Artifacts$, and $\Knowledge$ are durable record spaces; $\Gamma_t$ is the set of active P\&P leases; $\Runtime_t$ is volatile task state; $W$ is the non-agent Workflow Protocol; $G_t$ is the set of active Task Groups; and $H,C,T$ denote the human principal, Control Plane, and Translation Agent.

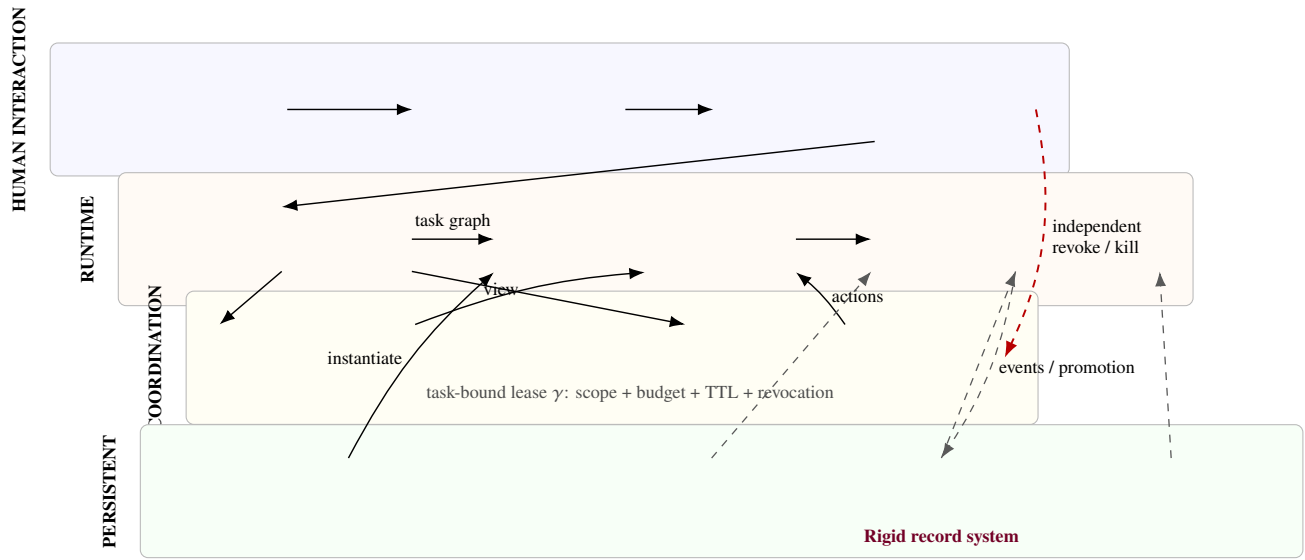
\begin{figure}[t]
  \centering
  \resizebox{0.98\textwidth}{!}{\begin{tikzpicture}[
  font=\small,
  component/.style={draw, rounded corners=2pt, minimum height=8.5mm, align=center, inner xsep=6pt, inner ysep=3pt},
  flow/.style={-{Latex[length=2.1mm]}, line width=.55pt},
  data/.style={-{Latex[length=2mm]}, dashed, line width=.5pt, draw=black!65},
  kill/.style={-{Latex[length=2.2mm]}, dashed, line width=.75pt, draw=red!72!black},
  band/.style={draw=black!28, rounded corners=3pt, inner sep=4.5mm},
  layerlabel/.style={font=\bfseries\scriptsize, rotate=90, anchor=south}
]
  \node[component, fill=blue!9] (human) at (0,0) {Human principal\\goal, veto, org kill};
  \node[component, fill=blue!9] (translator) at (4.5,0) {Translation agent\\intent $\rightarrow$ task input};
  \node[component, fill=blue!9] (control) at (9.3,0) {Control Plane\\OA/admin UI + minimal cache};
  \node[band, fit=(human)(translator)(control), fill=blue!3, on background layer] (humanband) {};
  \node[layerlabel] at ($(humanband.west)+(-2mm,0)$) {HUMAN INTERACTION};

  \node[component, fill=orange!13] (protocol) at (1.3,-1.75) {Workflow Protocol\\select, assemble, initiate};
  \node[component, fill=orange!13] (group) at (6.2,-1.75) {Dynamic Task Group\\temporary specialist workers};
  \node[component, fill=orange!13] (runtime) at (11.2,-1.75) {Runtime Store\\task cards, cache, heartbeat};
  \node[band, fit=(protocol)(group)(runtime), fill=orange!4, on background layer] (runtimeband) {};
  \node[layerlabel] at ($(runtimeband.west)+(-2mm,0)$) {RUNTIME};

  \node[component, fill=yellow!18] (permission) at (3.1,-3.35) {\textbf{Permission}\\compiled view of data, tools, channels};
  \node[component, fill=yellow!18] (privilege) at (8.9,-3.35) {\textbf{Privilege}\\typed authority to change state};
  \node[band, fit=(permission)(privilege), fill=yellow!5, on background layer] (coordband) {};
  \node[layerlabel] at ($(coordband.west)+(-2mm,0)$) {COORDINATION};
  \node[font=\scriptsize, text=black!70] at (6,-3.84) {task-bound lease $\gamma$: scope + budget + TTL + revocation};

  \node[component, fill=green!11, minimum width=45mm] (pool) at (2.2,-5.15)
    {Specialization + Role-Group Pool\\resident templates / team leads};
  \node[component, fill=purple!11] (log) at (7.1,-5.15) {Log\\append-only events};
  \node[component, fill=purple!11] (artifact) at (10.2,-5.15) {Artifacts\\versioned payloads};
  \node[component, fill=purple!11] (knowledge) at (13.3,-5.15) {Knowledge\\reviewed patterns};
  \node[band, fit=(pool)(log)(artifact)(knowledge), fill=green!3, on background layer] (persistband) {};
  \node[layerlabel] at ($(persistband.west)+(-2mm,0)$) {PERSISTENT};
  \node[font=\bfseries\scriptsize, text=purple!60!black] at (10.2,-5.78) {Rigid record system};

  \draw[flow] (human) -- (translator);
  \draw[flow] (translator) -- (control);
  \draw[flow] (control.south) -- (protocol.north);
  \draw[flow] (protocol) -- node[above, font=\scriptsize]{task graph} (group);
  \draw[flow] (group) -- (runtime);
  \draw[flow] (pool.north) to[bend left=10] node[left, font=\scriptsize]{instantiate} (group.south west);
  \draw[flow] (protocol.south) -- (permission.north west);
  \draw[flow] (protocol.south east) -- (privilege.north west);
  \draw[flow] (permission.north) to[bend left=8] node[left, font=\scriptsize]{view} (group.south);
  \draw[flow] (privilege.north) to[bend right=8] node[right, font=\scriptsize]{actions} (group.south east);
  \draw[data] (log.north) -- (runtime.south west);
  \draw[data] (artifact.north) -- (runtime.south);
  \draw[data] (knowledge.north) -- (runtime.south east);
  \draw[data] (runtime.south) to[bend left=12] node[right, font=\scriptsize]{events / promotion} (artifact.north);
  \draw[kill] (control.east) to[bend left=18] node[right, font=\scriptsize, align=left]{independent\\revoke / kill} (privilege.east);
\end{tikzpicture}}
  \caption{The four-layer architecture. Two persistent modules---the Specialization/Role-Group Agent Pool and the rigid record system---are coupled to a temporary Task Group through P\&P. The Workflow Protocol compiles the task, topology, working set, and lease. The Control Plane keeps a minimal independent cache so revocation and shutdown do not depend on the record data plane.}
  \label{fig:layers}
\end{figure}

The architecture deliberately separates persistence from activation. The persistent layer answers \emph{what reusable capabilities and organizational memory exist?} The coordination layer answers \emph{what world and what state changes are available to this identity now?} The runtime layer answers \emph{who performs this task, in what dependency shape, against which working set?} The human layer answers \emph{how does a principal express intent, inspect state, correct decisions, and stop the system?}

The end-to-end path is concise. A human event enters the Control Plane. The Translation Agent converts intent into typed input without deciding policy. The Workflow Protocol resolves dependencies, selects resident specializations, chooses a topology, instantiates workers, caches approved material into an isolated runtime, and issues P\&P leases. Work proceeds through task cards and records. Review or supervision may cause P\&P to be recompiled. Completion promotes approved outputs, appends state changes, distills reviewed knowledge, tears down the runtime, and revokes every lease. Fluidity is therefore the repeated compilation of $G_t$, $\Runtime_t$, $\Permission_t$, and $\Privilege_t$; rigidity lies in the schemas, write rules, identity separation, and accepted event history.

\subsection{Persistent layer: two modules}

\paragraph{Specialization/Role-Group Agent Pool.}
A resident entry is a reusable specification, not a permanently busy digital employee. It binds a domain specialization---for example evidence retrieval, financial modeling, interface design, or provenance checking---to one of the Role Groups in \Cref{sec:roles}. Its configuration may contain tools, skills, a working style or ``soul'' specification, reference packs, and default P\&P templates. It waits dormant in the pool. At activation, it acts as a team lead or creates temporary workers; those workers receive only the attenuated subset required for the current task. Because a new specialization can be added without changing the constitutional layers, the organization can extend its application repertoire without redesigning its foundation.

\paragraph{Four-store record system.}
The record system supports five organizational properties: accountability, traceability, restorability, reuse and improvement, and context management. ``Document system'' is too undifferentiated for this burden. The framework separates four stores by lifecycle and consistency requirement, as shown in \Cref{tab:stores}. Runtime is logically one of the four stores but physically belongs to the runtime layer; Log, Artifacts, and Knowledge form the durable part of the persistent layer.

\begin{table}[t]
\centering
\caption{Four logical stores and asymmetric write rules. The log is canonical for accepted state transitions; the Artifact Store is canonical for payload bytes.}
\label{tab:stores}
\footnotesize
\begin{tabularx}{\textwidth}{@{}p{2.05cm}p{2.55cm}YYp{2.45cm}@{}}
\toprule
Store & Contents & Write rule & Read/exposure rule & Failure role \\
\midrule
Runtime $\Runtime$ & task cards, approved cache, working payloads, grants, heartbeat, board state & operators write only here; typed state transitions & visible only inside the Task Group lease & disposable sandbox; rebuild or re-cache \\
Log $\Events$ & accepted events, authority use, lineage, handoff and decision records & kernel appends; no agent edits or deletes & sliced by Permission; full audit view is exceptional & replay state and reconstruct accountability \\
Artifacts $\Artifacts$ & approved deliverables, datasets, design records, versioned payloads & reviewer-authorized promotion with lineage and content hash & task-linked read, default current version & preserve production bytes; independent backup \\
Knowledge $\Knowledge$ & reviewed patterns, references, workflows, skills, incident lessons & approved change proposal only & progressive disclosure by task and specialization & reusable semantic/procedural memory \\
\bottomrule
\end{tabularx}
\end{table}

The asymmetry is a safety mechanism. An operator may create candidate artifacts in its runtime, but cannot silently promote them into the durable Artifact or Knowledge stores. A valid action automatically produces a log event, but agents do not receive a general ``edit history'' tool. A reviewer may authorize promotion only inside an activated decision scope. A supervisor may inspect broad slices and write evidence or FLAG records to runtime, but cannot rewrite the object under inspection. These store rules are Role-Group P\&P made concrete.

The log records observable organizational facts---actor, task, lease, event type, policy version, time, input and output hashes, claims, and evidence pointers---rather than private chain-of-thought. Work is considered organizationally complete only when its material result and authority exercise are represented by valid records. Direct messages can still wake an agent or point to a changed dependency, but specifications, evidence, approvals, and outputs do not live only in session history. This applies the end-to-end principle to organizational work: a delivery channel is an optimization; the durable correctness condition belongs in the record endpoint \citep{saltzer1984end}.

\paragraph{Archival and the Registrar.}
Archiving excludes obsolete versions from default retrieval; it does not delete their existence. Each document family has one current pointer, while historical versions require explicit opt-in. Immutable log segments can be checkpointed and moved to lower-cost retention without editing the accepted history in place. A narrow Registrar may maintain manifests, current pointers, checksums, indexes, and compaction schedules. It is an Operation specialization---a librarian, not an editor. It has no content authority; an inconsistency produces a FLAG rather than a repair made on its own judgment.

\paragraph{Recovery semantics.}
The phrase ``single source of truth'' needs a type. The Log is the source of truth for accepted state changes; the Artifact Store is the source of truth for payload bytes. If Runtime is lost, deterministic replay reconstructs task state from the event prefix, provided referenced artifacts remain reachable. A hash in an event cannot recreate deleted bytes, so artifacts require independent replication. If the Log is unavailable, Artifacts and Knowledge preserve approved production material and Runtime preserves in-flight work, but provenance and exact state history are degraded. This cross-store survival is useful, but it is not equivalent to complete recovery. The stores should therefore fail independently, with separate recovery-point and recovery-time objectives; the Log receives append-only, tamper-evident protection while Runtime is treated as replaceable.

\begin{proposition}[Scoped runtime recovery]
Let $P$ be a deterministic projector from a valid event prefix $\Events_{\le c}$ to runtime state, and assume every referenced Artifact payload remains reachable. If Runtime is lost, replay reconstructs $S_c=P(\Events_{\le c})$ and its artifact indexes. Replay does not reconstruct unavailable payload bytes.
\end{proposition}

This type-split preserves the v4 design intent---the record foundation allows work to be rebuilt---without assigning impossible recovery power to an event hash. Write-ahead logging and event replay provide mature engineering analogues \citep{mohan1992aries}; the organizational contribution is that authority, evidence, and work state share the replayable vocabulary.

\subsection{Coordination layer: Permission and Privilege}

P\&P is the architecture's most frequently changed coordination variable. It links the persistent modules to each transient execution and must therefore be called back in every layer.

\begin{definition}[Permission]
For agent $a$ at time $t$, $\Permission_a(t)$ is the externally enforced boundary of resources it may read, tools and channels it may invoke, and regions it may write. Permission defines the agent's observable and actionable world.
\end{definition}

\begin{definition}[Privilege]
For agent $a$ at time $t$, $\Privilege_a(t)$ is the set of organizational state changes its decision may cause: for example approve, reject, promote, publish, spawn, grant, revoke, adjust, or HALT.
\end{definition}

The distinction matters. Permission can be denied before inference by withholding a file handle, credential, tool, or channel. Privilege is procedural: exercising it necessarily needs enough Permission to inspect a case, so high privilege cannot be made safe merely by hiding every input. It is constrained through scope, expiry, countersignature, audit, reversibility, and human correction. Capability-oriented access control and separation-of-duty standards supply mature prior art for these controls \citep{miller2006robust,nist2020sp80053}; NIST has also identified identity, authorization, auditing, and non-repudiation as live requirements for software and AI agents \citep{nist2026identity}.

For task $\tau$, the harness compiles the agent's working view:
\begin{equation}
\mathcal{V}_a^\tau(t)=\operatorname{Compile}
\big(\Permission_a^\tau(t),\;I_\tau,\Runtime,\Artifacts,\Knowledge,\Events,\Pi\big).
\label{eq:view}
\end{equation}
The objective is not simply ``less context.'' It is a clean, authorized world: broad enough for the task, narrow enough to exclude unrelated or dangerous state. CoALA and MemGPT treat memory selection and paging as architectural concerns \citep{sumers2023coala,packer2023memgpt}; Equation~\eqref{eq:view} extends that idea to organizational data, tools, channels, and authority.

P\&P is issued as one task lease:
\begin{equation}
\gamma_a^\tau=\langle a,\tau,\Permission,\Privilege,b,t_0,t_{\mathrm{exp}},q\rangle,
\label{eq:lease}
\end{equation}
where $b$ is budget and $q$ a revocation handle. The lease is deny-by-default, task-bound, time-limited, and non-transferable except through an explicitly attenuating spawn operation. Grant and revocation are both logged. Completion, timeout, missed heartbeat, budget break, or HALT closes the lease; an in-flight tool call follows a predefined finish-or-cancel rule rather than leaving hanging authority.

A proposed event $e$ changes organizational state only if
\begin{equation}
\operatorname{Valid}(e,S_t)=\operatorname{Schema}(e)\land
\operatorname{LeaseActive}(e,\Gamma_t)\land
\operatorname{Authorized}(e,\Permission_t,\Privilege_t)\land
\operatorname{Precondition}(e,S_t).
\label{eq:valid}
\end{equation}
If valid, $S_{t+1}=\delta(S_t,e)$ and $e$ is appended to $\Events$. Natural language may propose $e$; the non-agent kernel commits it.

\begin{proposition}[Fail-closed authority]
Assume every material action is mediated by Equation~\eqref{eq:valid}, and expired or revoked leases cannot be forged. Stopping new lease issuance and revoking active leases prevents subsequent material state changes by agents.
\end{proposition}

The assumption is strong and visible. A side channel, shared credential, or tool that bypasses the harness invalidates the property. The framework does not call policy in a prompt ``enforcement.''
\subsection{Runtime layer: Workflow Protocol, Runtime Store, and Task Group}

The Workflow Protocol is an external system module under the Control Plane, not an agent with residual discretion. Its role is intentionally limited: translate a validated task input into a task graph; select specializations and a topology; initialize the runtime working set; issue and revoke P\&P; expose a lightweight board; and execute circuit breakers or kill commands. It supplies a starting workflow---who acts, on what, after which dependency, and under what completion condition---without pretending to prescribe every reasoning step.

A typed task card is
\begin{equation}
\tau=\langle id,g,I,O,\kappa,D,b,d,r,\ell\rangle,
\label{eq:task}
\end{equation}
where $g$ is the goal, $I$ approved inputs, $O$ required outputs, $\kappa$ completion and evaluation criteria, $D$ dependencies, $b$ budget, $d$ deadline, $r$ risk class, and $\ell$ lineage pointers. The card states what must become true and which evidence is admissible; it does not carry a full reasoning trace.

Given a task dependency graph $D_\tau$ and measured features $x_\tau$ such as coupling, uncertainty, tool density, reversibility, and risk, the protocol selects
\begin{equation}
\phi_\tau=\operatorname{Topology}(D_\tau,x_\tau)
\in\{\text{single},\text{parallel},\text{centralized},\text{decentralized},\text{hybrid}\},
\label{eq:topology}
\end{equation}
and assembles
\begin{equation}
G_\tau(t)=\operatorname{Assemble}\big(\tau,\Agents^{\mathrm{res}},\phi_\tau,\Pi\big).
\label{eq:assemble}
\end{equation}
Sequential dependency usually preserves a single chain; separable work permits parallel specialists; coupled synthesis introduces explicit aggregation points. Each aggregator has a fan-out parameter $f_{\max}$ because its context bandwidth is a designed resource. This is the concrete meaning of a fluid structure: $G_\tau$, $\phi_\tau$, worker instances, tools, readable documents, and privileges may all change when an event changes the task. Only specialization and Role Group remain comparatively stable.

The selection objective is net organizational utility, not maximum agent count:
\begin{equation}
\begin{split}
(G^*,\phi^*)=\arg\max_{G,\phi}\; &V_\tau\Pr(\mathrm{success}\mid G,\phi,\tau)
-C_{\mathrm{model}}-C_{\mathrm{tool}}\\
&-C_{\mathrm{coord}}-\lambda_L L-C_H-\mathbb{E}[\mathcal{L}_{\mathrm{risk}}].
\end{split}
\label{eq:utility}
\end{equation}
Human review $C_H$, coordination cost $C_{\mathrm{coord}}$, and expected loss are explicit. On a simple or highly sequential task, the optimizer should often choose one agent. Scaling evidence supports making topology conditional rather than constitutional \citep{kim2025scaling}.

\paragraph{Claim is commitment.}
Eligible cards appear on a board filtered by specialization and Permission. Claiming a card consumes a work-in-progress slot, starts a heartbeat, activates the lease, and records a budget and deadline commitment. Abandonment or missed heartbeat returns the card with an event. Completion delivers
\begin{equation}
h=\langle c,e,\hat p,k\rangle,
\label{eq:handoff}
\end{equation}
where $c$ is the claim, $e$ evidence, $\hat p$ confidence or uncertainty, and $k$ measured cost. Confidence is not trusted merely because a model states it; its calibration is an evaluation target. The tuple distinguishes ``submitted'' from ``believed correct'' and gives the next node an inspectable object.

\paragraph{Records carry; messages point.}
Agent-to-agent messages remain useful as interrupts: a dependency changed, a review is ready, or an artifact pointer should be opened. They are not the canonical work object. The protocol changes document visibility and task status; the next agent receives an authorized record set rather than an interpretation of another session. This reduces serial translation and makes a failed worker replaceable without losing the work substrate. It also improves asynchronous execution: an agent need not remain alive merely to explain its output.

\paragraph{Objective handoff.}
Where downstream persuasive intent could bias upstream evidence collection, a card withholds that preference and specifies only the deliverable standard. A source collector need not know which conclusion a later memo hopes to support. Review likewise starts from a clean context. This is not universal: jurisdiction, safety requirements, audience accessibility, and interface compatibility may define correctness and must remain visible. Objective handoff is therefore a conditional mechanism to test, not a doctrine that context is always harmful.

\subsection{Human-interaction layer: Control Plane and Translation Agent}

The Control Plane is the organization's OA/admin interface. It gives the human principal one surface for task initiation, board inspection, P\&P state, decision queues, FLAGs, budgets, and shutdown. It owns a minimal independent cache containing current active leases, policy versions, typed command schemas, and the stop switch. That cache is periodically recorded but is not dependent on the record stores for immediate revocation. Data-plane failure must not remove the brake.

The Translation Agent is a secretary and interpreter, not a machine CEO. Its sole function is to convert human language into proposed task cards, surface ambiguities, and render machine state back into human-readable views. It receives an Operation-class lease confined to the interface and cannot approve its own interpretation, allocate residual authority, or change policy. The human remains the correction endpoint, organization-kill holder, and one key for Major or Emergency changes. During human-offline periods, only pre-approved, reversible Standard work proceeds automatically; higher-risk decisions queue or HALT.

\section{Three Role Groups and Their Powers}
\label{sec:roles}

Specialization and Role Group answer different questions. Specialization asks \emph{what capability is useful?} Role Group asks \emph{what relationship to organizational truth and change is safe?} A marketing analyst may be instantiated as an Operator, a marketing-output Reviewer, or an independent Supervisor; the domain does not confer the power. The three groups are fixed constitutional categories even though every Task Group is temporary.

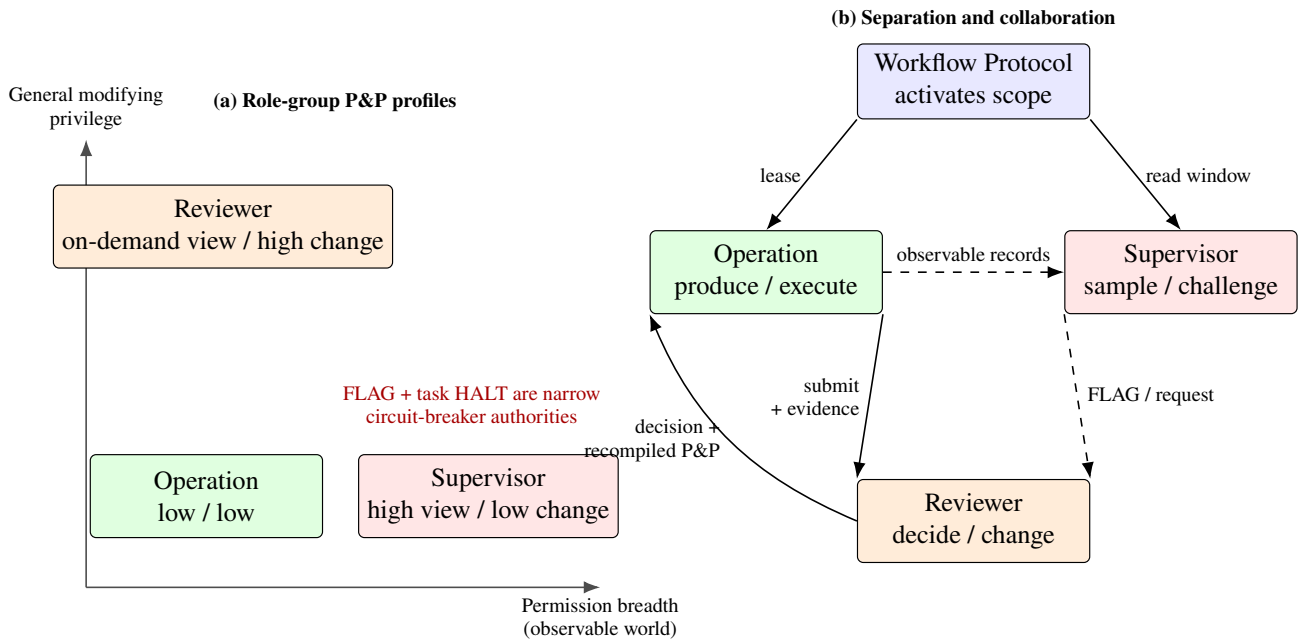
\begin{figure}[t]
  \centering
  \resizebox{0.98\textwidth}{!}{\begin{tikzpicture}[
  font=\small,
  role/.style={draw, rounded corners=2pt, minimum width=28mm, minimum height=10mm, align=center, inner sep=3pt},
  flow/.style={-{Latex[length=2mm]}, line width=.55pt},
  signal/.style={-{Latex[length=2mm]}, dashed, line width=.55pt},
  axis/.style={-{Latex[length=2mm]}, line width=.55pt, draw=black!70}
]
  \begin{scope}
    \draw[axis] (0,0) -- (6.2,0) node[below, align=center, font=\scriptsize] {Permission breadth\\(observable world)};
    \draw[axis] (0,0) -- (0,5.4) node[above, align=center, font=\scriptsize] {General modifying\\privilege};
    \node[role, fill=green!12] (opm) at (1.45,1.1) {Operation\\low / low};
    \node[role, fill=red!10] (supm) at (4.85,1.1) {Supervisor\\high view / low change};
    \node[role, fill=orange!15] (revm) at (1.65,4.35) {Reviewer\\on-demand view / high change};
    \node[font=\scriptsize, text=red!65!black, align=center] at (4.65,2.2)
      {FLAG + task HALT are narrow\\circuit-breaker authorities};
    \node[font=\bfseries\scriptsize] at (3.0,5.85) {(a) Role-group P\&P profiles};
  \end{scope}

  \begin{scope}[xshift=8.2cm]
    \node[role, fill=green!12] (op) at (0,3.8) {Operation\\produce / execute};
    \node[role, fill=red!10] (sup) at (5.0,3.8) {Supervisor\\sample / challenge};
    \node[role, fill=orange!15] (rev) at (2.5,0.8) {Reviewer\\decide / change};
    \node[draw, rounded corners=2pt, fill=blue!9, minimum width=28mm, align=center] (wp) at (2.5,6.1)
      {Workflow Protocol\\activates scope};
    \draw[flow] (wp.south west) -- node[left, font=\scriptsize]{lease} (op.north);
    \draw[flow] (wp.south east) -- node[right, font=\scriptsize]{read window} (sup.north);
    \draw[flow] (op.south east) -- node[left, font=\scriptsize, align=right]{submit\\+ evidence} (rev.north west);
    \draw[signal] (sup.south west) -- node[right, font=\scriptsize]{FLAG / request} (rev.north east);
    \draw[flow] (rev.west) to[bend left=22] node[left, font=\scriptsize, align=right]{decision +\\recompiled P\&P} (op.south west);
    \draw[signal] (op.east) -- node[above, font=\scriptsize]{observable records} (sup.west);
    \node[font=\bfseries\scriptsize] at (2.5,6.85) {(b) Separation and collaboration};
  \end{scope}
\end{tikzpicture}}
  \caption{Role Group is an asymmetric P\&P profile, not a title. Supervisor's broad observational Permission is paired with low modifying Privilege; FLAG and task HALT are narrow circuit-breaker authorities. Reviewer has high change Privilege but no standing panoramic view: the Workflow Protocol activates a decision-specific read window.}
  \label{fig:roles}
\end{figure}

\begin{table}[t]
\centering
\caption{The three Role Groups. ``High Permission'' for Supervision means broad potential coverage through sampled, compiled views, not that every record is loaded into one context.}
\label{tab:roles}
\scriptsize
\begin{tabularx}{\textwidth}{@{}p{1.55cm}p{2.15cm}p{2.35cm}Y Y p{2.5cm}@{}}
\toprule
Group & Permission profile & Privilege profile & Normal work & Writes & Cannot do \\
\midrule
Operation & low organization-wide; narrow task runtime plus approved read slices & low; tool execution and bounded spawn inside lease & produce, transform, call tools, submit evidence & Runtime only; system appends events; durable promotion is proposed & change P\&P or policy, edit other groups, approve own output, write durable stores directly \\
Reviewer & low standing visibility; decision-specific view activated externally & high scoped change authority: approve, reject, promote, adjust, roll back, grant/revoke & test request, evidence, criteria, reversibility, and blast radius; decide & typed decisions may change all stores and active leases within scope & self-activate a case, self-expand scope, bypass countersignature, erase history \\
Supervisor & high observational coverage across four stores; excludes private reasoning chain & low general modification; request status, FLAG, and narrow task HALT & sample work, run canaries, test provenance, adversarially challenge, monitor missing records & Runtime evidence, monitoring state, FLAG/HALT event only & approve, promote, rewrite artifact/knowledge/log, distribute P\&P, impose preferred substitute \\
\bottomrule
\end{tabularx}
\end{table}

\paragraph{Operation: low Permission, low Privilege.}
Operation performs the work. Its macroscopic world is intentionally narrow: one runtime sandbox, approved input slices, named tools, and a bounded output location. It may read task-relevant Artifact, Knowledge, or Log slices but writes only Runtime. A resident operator may instantiate workers under a pre-authorized template, yet every child receives attenuated grants and remains low/low. Neither local competence nor group-lead status creates organizational authority.

\paragraph{Reviewer: high Privilege, on-demand Permission.}
Review owns decision and adjustment. It can approve or reject work, promote candidate payloads, authorize durable Knowledge, change a workflow, issue a new lease, roll back a reversible transition, or request a fresh Task Group. Because these are high privileges, Reviewer is dormant when no decision is pending. Permission is activated by the Workflow Protocol only after a task-card submission or evidence-bearing Supervisor request. The Reviewer first validates the request and its evidence, then validates the proposed action; it cannot trigger its own broad read window. Major or irreversible change requires a second key from a distinct Reviewer, a deterministic control, or the human.

\paragraph{Supervisor: high observational Permission, low modifying Privilege.}
Supervision is designed from a falsification stance: assume an Operation may not be working correctly, then try to reject that hypothesis with evidence. It samples all four stores, requests status, checks that actions landed in the record, runs known-defect canaries, and may independently construct a clean-context comparison plan. It examines observable outputs and actions rather than inheriting the producer's reasoning narrative. A FLAG must carry a replayable evidence pointer. Supervisor cannot approve its preferred replacement, promote an artifact, change P\&P, or rewrite policy. Its ability to HALT a task is a separate circuit-breaker authority: stopping propagation does not grant the general right to decide what should replace the stopped work.

The collaboration loop makes the separation operational. Operation submits work and evidence. Supervisor observes the record and, when warranted, sends a FLAG or adjustment request to Reviewer. Reviewer is activated for that case, double-checks both request and proposed response, and commits or rejects a typed decision. The Workflow Protocol then recompiles P\&P and runtime state; Operation resumes under a new lease. Operation cannot approve, Supervisor cannot alter, and Reviewer cannot self-activate. No agent controls the full chain.

This separation is logical before it is physical. In a small, low-risk prototype, one model may perform Review and Supervision in clean, sequential contexts to control coordination cost. For consequential work, identities and contexts must be distinct; where correlated model blind spots matter, model families or deterministic checks should also differ. Information separation alone does not guarantee epistemic independence, but it prevents the most direct form of self-validation. Artificial Organisations provides a close empirical neighbor by enforcing compartmentalized corroboration and critique \citep{waites2026artificial}.
\section{Why the Integrated Design Is Different}

Classical MAS already distinguishes an agent, its environment, commitments, and protocols \citep{wooldridge2009mas}; Contract Net formalized announcement, bidding, and award decades before LLM role prompts \citep{smith1980contract}. Recent criticism is therefore right that a dialogue among LLMs should not automatically be treated as a mature multi-agent organization \citep{lamalfa2025mark}. The present framework's claim is narrower than ``first use of roles,'' ``first dynamic team,'' or ``first shared memory.'' Its claimed contribution is an integrated control path:
\begin{equation}
\boxed{
\text{Persistent Pool + Rigid Record}
\xrightarrow[\text{task/event}]{\text{Workflow Protocol + P\&P}}
\text{Dynamic Task Group}
}
\label{eq:corepath}
\end{equation}
P\&P is the coupling mechanism, not an isolated security feature. Permission selects the record slices, tools, channels, and runtime that constitute an agent's world. Privilege determines which task, record, lease, or organizational state transitions its decision may commit. The same variables drive team activation, document handoff, review visibility, supervisory observation, recovery scope, and shutdown.

\begin{table}[t]
\centering
\caption{Mechanism-level comparison with representative approaches. ``Partial'' recognizes overlap; the contribution is the joined control path in Equation~\eqref{eq:corepath}, not ownership of every component.}
\label{tab:comparison}
\scriptsize
\begin{tabularx}{\textwidth}{@{}p{2.45cm}p{2.15cm}p{2.35cm}p{2.25cm}Yp{2.45cm}@{}}
\toprule
Approach & Work structure & Canonical handoff & Main control variable & Governance & Failure substrate \\
\midrule
Conversational roles (CAMEL, AutoGen) \citep{li2023camel,wu2023autogen} & prompted roles and dialogue topology & conversation / message & prompt and speaker policy & optional evaluator or human & conversation retry; framework-dependent state \\
Fixed specialist workflow (MetaGPT) \citep{hong2023metagpt} & role/SOP pipeline & messages plus produced documents & workflow and role prompt & staged checks & workflow-local artifacts and retry \\
Dynamic organization layer (OMC) \citep{yu2026skills} & recruited Talents and execution tree & typed interfaces, tasks, meetings & talent/container lifecycle and tree search & explore--execute--review & retry and escalation through interfaces \\
Compartmentalized verification \citep{waites2026artificial} & fixed composition roles & composition artifacts & enforced information asymmetry & corroborator + critic & iterative correction and refusal behavior \\
This framework & event-assembled specialization topology & typed cards, artifacts, evidence, lineage & Permission + Privilege leases & Operation / Review / Supervision with human endpoint & four stores, replay, quarantine, scoped restart, independent kill \\
\bottomrule
\end{tabularx}
\end{table}

Three consequences follow.

\paragraph{First, attention control becomes enforceable organization design.}
A prompt that says ``focus only on the relevant data'' leaves the noisy world present and asks a probabilistic model to ignore it. Permission removes unrelated resources from the compiled view. This does not guarantee correctness, but it makes context selection inspectable, reproducible, and revocable. The framework manages what the agent need not consider, which is an organizational information-processing function rather than prompt craftsmanship.

\paragraph{Second, communication becomes durable coordination.}
A message handoff requires the next agent to interpret a previous context. A record handoff gives it a typed task, payload, evidence, uncertainty, and lineage under a new Permission boundary. If one worker fails or disappears, the organization retains the work object. This directly targets context drift, serial translation loss, and irreversible session-bound failure without claiming to eliminate semantic misunderstanding.

\paragraph{Third, adaptability obtains a stable boundary.}
A Task Group can change composition and workflow in real time because its workers are temporary and its P\&P is leased. The record system and Role Groups stay rigid so dynamic assembly does not become ungoverned negotiation. This is neither a static org chart nor a fully flat swarm. It is a contingency structure: fluid execution compiled over a persistent frame.

The design is therefore \emph{management, not guidance}. It does not attempt to describe correct behavior in ever longer prompts. It establishes task objects, access boundaries, state-transition procedures, feedback paths, and recovery rules, then allows specialized agents discretion inside those constraints.

\section{Robustness, Escalation, and Control}

Robustness is the ability to absorb error without converting one bad output or failed component into organizational collapse. No evaluator, prompt, store, or agent is assumed perfect. Following defense-in-depth logic from high-reliability systems \citep{reason1997managing}, the framework stacks five different responses (Cref{fig:defense}).

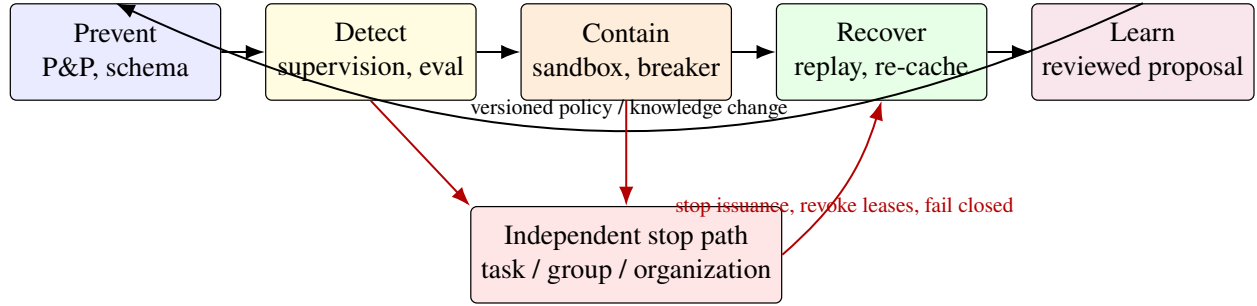
\begin{figure}[t]
  \centering
  \resizebox{0.94\textwidth}{!}{\begin{tikzpicture}[
  font=\small,
  stage/.style={draw, rounded corners=2pt, minimum width=24mm, minimum height=11mm, align=center},
  flow/.style={-{Latex[length=2.2mm]}, line width=.65pt},
  killbox/.style={draw, rounded corners=2pt, minimum width=31mm, minimum height=11mm, align=center, fill=red!10}
]
  \node[stage, fill=blue!8] (prevent) {Prevent\\P\&P, schema};
  \node[stage, fill=yellow!14, right=5mm of prevent] (detect) {Detect\\supervision, eval};
  \node[stage, fill=orange!14, right=5mm of detect] (contain) {Contain\\sandbox, breaker};
  \node[stage, fill=green!10, right=5mm of contain] (recover) {Recover\\replay, re-cache};
  \node[stage, fill=purple!10, right=5mm of recover] (learn) {Learn\\reviewed proposal};
  \draw[flow] (prevent) -- (detect);
  \draw[flow] (detect) -- (contain);
  \draw[flow] (contain) -- (recover);
  \draw[flow] (recover) -- (learn);
  \draw[flow] (learn.north) to[bend left=25] node[above, font=\scriptsize]{versioned policy / knowledge change} (prevent.north);

  \node[killbox, below=12mm of contain] (kill) {Independent stop path\\task / group / organization};
  \draw[flow, red!70!black] (detect.south) -- (kill.north west);
  \draw[flow, red!70!black] (contain.south) -- (kill.north);
  \draw[flow, red!70!black] (kill.east) to[bend right=16]
    node[below, font=\scriptsize]{stop issuance, revoke leases, fail closed} (recover.south);
\end{tikzpicture}}
  \caption{The governance loop. Prevention constrains reachable states; Supervision detects; isolation contains; record replay and clean context recover; Review converts incidents into versioned proposals. The independent stop path acts on leases rather than negotiating through the failed data plane.}
  \label{fig:defense}
\end{figure}

\subsection{Prevent, detect, contain, recover, learn}

\textbf{Prevent} combines deny-by-default P\&P, task-card schema validation, dependency-cycle checks, objective handoffs, budget ceilings, and write asymmetry. \textbf{Detect} combines heartbeat, acceptance tests, Supervisor sampling, canary tasks, integrity checks, and measured anomalies. \textbf{Contain} uses per-group runtime sandboxes, channel circuit breakers, and lineage quarantine: when an artifact is contaminated, descendants become unverified and stop propagating. \textbf{Recover} uses re-cache, deterministic replay, clean-context restart, task return to the board, and rollback where the action is reversible. \textbf{Learn} creates a reviewed incident proposal for Knowledge or policy; the incident never edits the constitution directly.

Runtime recovery is intentionally cheap. If a sandbox corrupts, Supervisor attaches evidence, Reviewer authorizes recovery, and the Workflow Protocol reconstructs approved inputs from Artifacts and Knowledge plus state from the Log. The group is re-instantiated with narrower P\&P. If a durable store fails, recovery follows the typed semantics in \Cref{tab:stores}; a surviving Log restores state references, while surviving Artifacts preserve production bytes. Control Plane revocation remains available in either case.

\subsection{Supervision tree and clean restart}

The instantiation hierarchy is also a supervision tree: temporary worker $\rightarrow$ resident specialization lead $\rightarrow$ review domain $\rightarrow$ human. Erlang/OTP demonstrates the engineering value of isolating process failure and restarting from known state \citep{armstrong2003reliable}. In an LLM context, clean restart has an additional rationale: it discards a confused or contaminated conversational state. Recovery loads the last accepted checkpoint, supplies external error evidence, and narrows the task. It does not merely ask the same context to ``reflect harder,'' because intrinsic self-correction is unreliable without trustworthy external feedback \citep{huang2024selfcorrect,kamoi2024selfcorrect}.

Restart scope follows dependency: one-for-one for an isolated worker, rest-for-one for its downstream lineage, and one-for-all only for inseparable shared state. A restart-intensity limit $(N,T)$ converts repeated failure into escalation. Values such as $N=3$ are prototype parameters, not universal laws; the mechanism is the limit and evidence-bearing handoff.

\subsection{Escalation, change classes, and kill scopes}

An error is first classified as \emph{transient}, \emph{deterministic}, or \emph{systemic}. A transient tool timeout may retry under the same card. A deterministic schema or permission failure should not repeat the same route; it BLOCKs and escalates. A systemic fault affecting multiple groups or the control path trips an organization breaker. Escalation reuses Equation~\eqref{eq:handoff}: claim, evidence, confidence, and cost. Its route is typed as ASSIGN, BLOCK, REVIEW, FLAG, or HALT.

Change authority is calibrated on two axes: reversibility and blast radius (\Cref{tab:classes}). This makes Reviewer power explicit rather than pretending that approval has no cost.

\begin{table}[t]
\centering
\caption{Risk-based decision classes. Thresholds are organization-specific and must be calibrated.}
\label{tab:classes}
\small
\begin{tabularx}{\textwidth}{@{}p{1.8cm}YYp{3.1cm}@{}}
\toprule
Class & Typical property & Required authority & Recovery / record requirement \\
\midrule
Standard & repeated, low blast radius, reversible & operator under pre-approved policy & automatic rollback; sampled later \\
Normal & material but reversible & scoped Reviewer decision & logged evidence and rollback plan \\
Major & high blast radius or difficult to reverse & Reviewer plus independent second key or human & checkpoint, explicit evidence, human-visible decision \\
Emergency & containment or recovery after incident & break-glass identity within predefined bounds & immediate record plus mandatory post-incident review \\
\bottomrule
\end{tabularx}
\end{table}

The stop system has three scopes. \textbf{Task kill}: an Operator may halt its own task, and Supervisor may halt any task as a narrow circuit breaker. \textbf{Group kill}: Reviewer authorizes shutdown, while an automatic safety tripwire or human may bypass a failed Reviewer. \textbf{Organization kill}: human-only, one action, always available---stop new leases and revoke every active lease. Circuit breakers and bulkheads apply familiar distributed-systems patterns \citep{nygard2007release}. Because agents are dormant by default, shutdown does not need to persuade or locate a permanently autonomous workforce; it stops the exceptional state of active authority.

\subsection{Measurement is part of control}

A framework justified by efficiency must meter its coordination. Each task records model and tool cost, wall time, retries, human minutes, evidence volume, P\&P changes, and incidents. Three diagnostic measures are
\begin{align}
\chi_{\mathrm{coord}} &= \frac{C_{\mathrm{routing}}+C_{\mathrm{handoff}}+C_{\mathrm{review}}+C_{\mathrm{supervision}}}
{C_{\mathrm{productive\ work}}},\\
A_{\mathrm{audit}} &= \frac{\#\text{ material transitions with valid lease and evidence}}
{\#\text{ material transitions}},\\
E_{\mathrm{amp}} &= \frac{\#\text{ invalid downstream artifacts}}
{\#\text{ injected or identified upstream faults}}.
\end{align}
The first exposes coordination tax; the second tests whether ``power is recorded'' is actually true; the third measures propagation. Budgets are enforced, not merely visualized. Task-, group-, and organization-level ceilings HALT execution, while an error budget freezes change work after a service objective is exhausted, following site-reliability practice \citep{beyer2016sre}. Metrics do not replace outcome quality, and they can be gamed; acceptance criteria and monitoring policy must therefore be versioned and, when appropriate, hidden from the producer's writable context.
\section{Status, Limitations, and Research Agenda}

\subsection{Prototype status and next validation}

This work is a design-science artifact: it specifies constructs, architecture, propositions, and a method of evaluation \citep{hevner2004design,gregor2007anatomy}. A prototype version has been implemented and exercised in small-sample tests. Those exercises support the claim that the design can be instantiated and have been useful for refining mechanics; they are not a large-scale, powered empirical evaluation and do not justify a general claim that the full organization outperforms alternative systems. Large-scale validation has not yet been completed.

The companion empirical paper should compare: (B0) a strong single agent; (B1) prompt-defined roles with conversational handoff; (B2) a fixed graph or SOP workflow; (B3) a recent organization-layer system such as OMC where reproducible; and (B4) the full architecture. Ablations should remove P\&P leasing, Role-Group separation, record handoff, four-store separation, topology selection, or replay one at a time. Tasks must vary by dependency shape, tool density, uncertainty, reversibility, and blast radius. MultiAgentBench and MAST provide useful topology and failure vocabularies \citep{zhu2025multiagentbench,cemri2025mast}, but consequential business cases also require blinded human evaluation because many public benchmarks have clear answers and symmetric losses.

The framework makes falsifiable predictions:
\begin{description}[style=nextline,leftmargin=1.05cm,labelwidth=.85cm]
  \item[H1] P\&P-compiled views and record handoffs reduce specification, handoff, and termination failures relative to prompt-only transfer under matched models and budgets.
  \item[H2] Task-conditioned topology improves net utility over a fixed multi-agent topology; on sequential tasks it frequently selects one chain.
  \item[H3] Task-leased P\&P reduces exposed action surface, stale authority, and incident blast radius without an unacceptable rate of legitimate denial.
  \item[H4] Operation/Review/Supervision separation improves defect detection as consequence rises, but can reduce net utility on routine work because validation is not free.
  \item[H5] Event replay plus replicated artifacts lowers recovery time and state divergence under injected runtime, index, context, and worker failures.
\end{description}
Primary outcomes should include task quality, Equation~\eqref{eq:utility}, human intervention minutes, expected-loss proxies in a sandbox, time to detect, recovery time, stale-grant duration, $\chi_{\mathrm{coord}}$, $A_{\mathrm{audit}}$, and $E_{\mathrm{amp}}$. Model family, tools, prompt, compute budget, and task order should be controlled or randomized; high-risk results should be repeated with heterogeneous evaluators.

\subsection{Limitations}

\paragraph{Mediation is an assumption, not magic.}
Fail-closed authority requires every material tool and credential to pass through the harness. Shared secrets, administrator mistakes, compromised control code, or unobserved side channels invalidate the guarantee.

\paragraph{P\&P can be wrong.}
A compiled view can omit validity-defining context, and a privilege template can deny legitimate work or expose too much. Least privilege is a calibration problem as well as a principle. False denials, emergency overrides, stale grants, and over-broad read windows must be measured.

\paragraph{Separation does not guarantee independence.}
Distinct contexts using the same model, sources, and rubric can share blind spots. Supervisor's adversarial prior can also create review load. Evidence requirements, canary tasks, heterogeneous models, deterministic tests, and human countersignature reduce but do not eliminate correlated error.

\paragraph{Records create cost and risk.}
Append-only events grow; artifacts need independent backup; logs can expose sensitive metadata. Retention classes, encryption, access review, tombstone-based redaction, verifiable compaction, and privacy threat modeling remain implementation work. The Registrar itself can become an indexing failure point even though it has no content authority.

\paragraph{Dynamic organization has coordination tax.}
Assembly, handoff, supervision, and review consume tokens, tools, latency, and human attention. Fixed thresholds for fan-out, heartbeat, review frequency, restart intensity, and error budgets would be false precision. They require domain-specific calibration, and a strong single agent may remain optimal for many tasks.

\paragraph{Human control limits unattended autonomy.}
The present target is a human-supervised organization. Offline policy deliberately queues or halts higher-risk decisions. This sacrifices 24-hour decision throughput to preserve a human correction endpoint; fully autonomous constitutional amendment is out of scope.

\subsection{Further research: the Agent OS analogy}

Only as a future research lens, the architecture resembles an operating system. The rigid organizational frame is a kernel; the specialization pool plus Artifact and Knowledge stores resemble applications and durable data; P\&P and the Workflow Protocol resemble controlled dependencies and system calls; Runtime resembles task-local working memory. This analogy suggests research on scheduling, isolation, portability, observability, and error budgets. It should not obscure the present paper's object: an organization-design framework, not a claim to have built a general agent operating system.

The planned sequence is therefore: (1) the present framework and logic; (2) controlled, larger-sample data on quality, cost, recovery, and risk; and (3) scale, nested organizations, domain-sharded Review, and carefully governed self-evolution. Organizational learning should remain proposal-based: agents may generate variants and evidence, but only reviewed transitions alter Knowledge or policy.

\section{Conclusion}

An agent organization should be designed around its substrate rather than dressed in human titles. Its members are replaceable instances with specialized capability and limited context; its durable identity lies in the record, policy, and reusable specialization templates. The appropriate architecture is consequently asymmetric.

The persistent foundation contains a rigid four-store record system and a dormant Specialization/Role-Group Agent Pool. The coordination layer uses Permission to compile the world an agent may observe and Privilege to constrain the state changes it may cause. The runtime layer lets the Workflow Protocol assemble a Task Group, cache an authorized working set, and issue expiring leases according to the current event and dependency shape. The human layer preserves intent, correction, and shutdown through an independent Control Plane and a translation-only interface agent.

Three Role Groups give the design its governance boundary: Operation is low/low and writes Runtime; Reviewer has high but on-demand change authority; Supervisor has broad observational Permission, little modifying Privilege, and a narrow stop circuit. Their collaboration makes P\&P more than access control: it becomes the organization's principal coordination instrument. The surface can therefore remain fluid while records, write rules, separation of powers, and the human endpoint remain rigid.

The framework is not presented as a completed empirical verdict. A prototype and small-sample exercises exist; large-scale validation is forthcoming. The paper offers a coherent object to implement, attack, measure, and revise: manage agents by compiling their world and authority, preserve work outside their sessions, and let dynamic organization occur only over a recoverable foundation.

\section*{AI Assistance Disclosure}
A prototype of the agent system described in this paper assisted with literature organization, drafting, and typesetting. The core design ideas, framework, and architecture were produced by the author, who remains responsible for all claims, source verification, authorship, and any submitted version.

\begingroup
\fontsize{7.35}{8.3}\selectfont
\setlength{\bibsep}{0pt plus .12pt}
\bibliographystyle{plainnat}
\bibliography{references}
\endgroup

\end{document}